\documentclass[12pt,a4paper,twoside]{article}

\usepackage[cp1251]{inputenc} 
\usepackage[T1]{fontenc} 
\usepackage[sc]{mathpazo} 
\usepackage{microtype} 
\usepackage{mathrsfs}
\usepackage[english]{babel} 
\usepackage{bbm}
\usepackage{amsmath,amsfonts,amscd,amssymb,latexsym}

\usepackage[unicode, dvipdfmx]{hyperref}
\usepackage[usenames]{color}

\usepackage{graphics}
\usepackage{titling}
\usepackage{ textcomp }
\graphicspath{{/}}
\DeclareGraphicsExtensions{.pdf,.png,.jpg}

\usepackage{amssymb}
\usepackage{amsmath}
\usepackage[dvips]{graphicx}
\usepackage{rotating}
\usepackage{longtable}
\usepackage{multirow}
\usepackage{geometry} 
\usepackage{fancyhdr}
\fancyheadoffset{0mm}
\fancyfootoffset{0mm}
\fancypagestyle{plain}{ 
    \fancyhf{}
    \rhead{\thepage}}
\begin{document}


\newenvironment{sciabstract}{
\begin{quote} \scriptsize}
{\end{quote}}

\pretitle{\begin{center}\Huge\bfseries} 

\title{
Fusion cross section and total kinetic energy of fission fragments by the dynamical dissipative surface-friction model}

\author
{S. Amano$^{1}$, Y. Aritomo$^{1}$, M.Ohta$^{2}$\\
\scriptsize{$^{1}$Kindai University Higashi-Osaka, Osaka 577-8502, Japan}\\
\scriptsize{$^{2}$Konan University Kobe, Hyogo 658-8501, Japan}\\
\scriptsize{e-mail: amano.shota3@gmail.com}\\
}

\date{}
\maketitle

\begin{flushleft}

\end{flushleft} 

\begin{sciabstract}

The capture cross section, the fusion cross section, and the quasi-fission yield producing symmetric fragments ($A_{CN}/2\pm20u$) in the $^{48}$Ca+$^{238}$U reaction are analyzed by the multidimensional Langevin equation taking account of the surface friction effect.
From the experimental data, the strength of the tangential friction has been determined.
It is presented that tangential friction increases in proportional to the power of the relative velocity of the projectile and the target.



\end{sciabstract}

\section*{Introduction}

Understanding the nuclear structure and fusion dynamics in low-energy heavy ion reactions is very important for proposing effective experimental methods and experimental setup.
In particular, in order to synthesize superheavy nuclei and to produce neutron-rich nuclei, the studies on the reaction dynamics is called for.
Recently, the 118 element has synthesized and the experimental trials for new superheavy elements have been continued.
Many attempts will pursuits the limit of existence of elements beyond the 118 element to show how nucleons can be bound to form mono nucleus.
In the r-process studies [1, 2, 3], the production of neutron-rich nuclei far from the stable line is crucial.
The knowledge for neutron-rich superheavy nuclei is also useful to get new insights into the nucleosynthesis and chemical evolution of the universe.
In order to produce superheavy nuclei, it is essential to understand the reaction dynamics and nuclear structure.
Here, the reaction dynamics is investigated by focusing on the friction between reaction partners in the fusion process.
In the first stage, when the projectile and the target nucleus start to come into contact with each other, the friction in the radial and the tangential direction take effect through the contact surface.
The radial friction causes the dissipation of their kinetic energy. On the other hand, the tangential one causes the dissipation of orbital angular momentum [4].
We focused on tangential friction in this paper.
There are two types of tangential friction: the sliding friction and the rolling friction. These are discussed theoretically by the surface-friction model [5], but the strength of the friction  is still under studies.
The coefficients of the friction are parameters to explain experimental data in the model but no comparison with the experimental data is yet shown [5].
The nature of the sliding friction is known to affect on the reaction process much more than the rolling friction.
The rotation of the contact system is controlled by the friction relating with the relative velocity between the projectile and the target.
It is reported that the rolling friction is shown to have less effect on the reaction cross section than the sliding friction [4].
In this paper, we investigate the dynamics of the reaction due to the sliding friction.
In the followings, we renamed the sliding friction as the tangential friction.
The friction in the tangential direction converts a part of the orbital angular momentum into the spin of the nucleus, and as the results the centrifugal potential energy decreases.
Less fusion reactions are observed in the reaction with large orbital angular momentum, but fusion reactions appear as decreasing the centrifugal potential energy.
Here, we clarified the dynamical property of the tangential friction by investigating in various incident energies.
The tangential friction in consistent with the experimental quantities in the fusion process, that is the complete fusion and the yield concerning with the symmetric fission, is shown.
Such study also leads to elucidate the mechanism of fusion process.

The framework of the present model is described briefly in the following section. The numerical results are compared with the experimental data to determine the strength of the tangential friction in the succeeding section. The final section is devoted to the summary and discussions.

\section*{Framework}
\subsection*{Potential energy}

\quad We adopt the dynamical model which is established as the unified model [6]. The initial stage of the nucleon transfer occurred in the reaction consists of two parts: (1) the system composed of the projectile and the target in their ground state and separating at infinity starts to reconfigure each single particle state.
(2) In the part where the projectile and the target fuse each other, the potential energy surface changes from the diabatic one to the adiabatic one.
Therefore, we consider the time evolution of potential energy from the diabatic one $V_{diab}\left(q\right)$ to adiabatic one $V_{adiab}\left(q\right)$.
Here, $q$ denotes a set of collective coordinates representing nuclear deformation. The diabatic potential is calculated by a folding procedure using effective nucleon-nucleon interaction [6, 7, 8].
The adiabatic potential energy of the system is calculated using an extended two-center shell model [8].
Then, we connect the diabatic and the adiabatic potentials with a time-dependent weighting function as follows:
\begin{eqnarray}
&&V=V_{diab}\left(q\right)f\left(t\right)+V_{adiab}\left(q\right)\left[1-f\left(t\right)\right], \nonumber \\
&&f\left(t\right)=\exp{\left(-\frac{t}{\tau}\right)}.
\label{pot}
\end{eqnarray}
Where $t$ is the interaction time and $f\left(t\right)$ is the weighting function with the relaxation time $\tau$.
We use the relaxation time $\tau=10^{-21}s$ proposed in [9, 10, 11].
We use the two-center parameterization [12, 13] to coordinated the nuclear deformation.
To solve the dynamical equation numerically and avoid the huge computation time,
we strictly limited the number of degrees of freedom and employ three parameters as follows:
$z_{0}$ (distance between the centers of two potentials),
$\delta$ (deformation of fragment), and $\alpha$ (mass asymmetry of colliding nuclei);
$\alpha=\frac{\left(A_{1}-A_{2}\right)}{\left(A_{1}+A_{2}\right)}$,
where $A_{1}$ and $A_{2}$ stand for the mass numbers of the target and projectile,
respectively [6, 14] and are used also as the mass numbers of the two fissioning fragments.
As shown in Fig.~1 in Ref. [12], the parameter $\delta$ is defined as $\delta=\frac{3\left(a-b\right)}{\left(2a+b\right)}$, where $a$ and $b$ represent the half  length of the long and the short elliptic axes in the $z_{0}-\delta$ space, respectively.
We assume that each fragment has the same deformation.
In addition, we use scaling technics to save computation time and use the coordinate $z$ defined as $z=\frac{z_{0}}{\left(R_{CN}B\right)}$, where $R_{CN}$ denotes the radius of the spherical compound nucleus and the parameter $B$ is defined as $B=\frac{\left(3+\delta\right)}{\left(3-2\delta\right)}$.
\subsection*{Multidimensional Langevin equation}

\quad We perform trajectory calculations of the time-dependent unified potential energy [6, 7, 14] by Langevin equation.
We start trajectory calculations from a sufficiently long distance between both nuclei [14].
So, we use the model which takes into account the nucleon transfer for slightly separated nuclei [6].
For the nucleon transfer between two separate nuclei use is made of the procedure described in Refs. [6, 7].
When both nuclei fuse into the mononucleus having sufficient wide neck, the evolution process of the mass asymmetry parameter $\alpha$ switches from the master equation to Langevin equation according to the procedure described in Ref. [14].
We use the multidimensional Langevin equation [6, 14, 15] as follows:
\begin{eqnarray}
&\frac{dq_{i}}{dt}=\left(m^{-1}\right)_{ij}p_{j}, \nonumber \\
&\frac{dp_{i}}{dt}=-\frac{\partial V}{\partial q_{i}}-\frac{1}{2}\frac{\partial}{\partial q_{i}}\left(m^{-1}\right)_{jk}p_{j}p_{k}
\gamma~-_{ij}\left(m^{-1}\right)_{jk}p_{k}+g_{ij}R_{j}\left(t\right), \nonumber \\
&\frac{d\theta}{dt}=\frac{\ell}{\mu_{R}R^{2}}, \nonumber \\
&\frac{d\varphi_{1}}{dt}=\frac{L_{1}}{\Im_{1}}, \nonumber \\
&\frac{d\varphi_{2}}{dt}=\frac{L_{2}}{\Im_{2}}, \nonumber \\
&\frac{d\ell}{dt}=-\frac{\partial V}{\partial\theta}-\gamma_{tan}\left(\frac{\ell}{\mu_{R}R^{2}}-\frac{L_{1}}{\Im_{1}}a_{1}-\frac{L_{2}}{\Im_{2}}a_{2}\right)R
+Rg_{tan}R_{tan}\left(t\right), \nonumber \\
&\frac{dL_{1}}{dt}=-\frac{\partial V}{\partial\varphi_{1}}-\gamma_{tan}\left(\frac{\ell}{\mu_{R}R^{2}}-\frac{L_{1}}{\Im_{1}}a_{1}-\frac{L_{2}}{\Im_{2}}a_{2}\right)a_{1}
-a_{1}g_{tan}R_{tan}\left(t\right), \nonumber \\
&\frac{dL_{2}}{dt}=-\frac{\partial V}{\partial\varphi_{2}}+\gamma_{tan}\left(\frac{\ell}{\mu_{R}R^{2}}-\frac{L_{1}}{\Im_{1}}a_{1}-\frac{L_{2}}{\Im_{2}}a_{2}\right)a_{2}
-a_{2}g_{tan}R_{tan}\left(t\right).
\label{lan}
\end{eqnarray}
The collective coordinates $q_{i}$ represent $z, \delta$, and $\alpha,$ the symbol $p_{i}$ denotes  momentum conjugated to $q_{i}$, and $V$ is the multidimensional potential energy. The symbols $\theta$ and $\ell$ indicates the relative  orientation of nuclei and relative angular momentum respectively. $\varphi_{1}$ and $\varphi_{2}$ stand for the rotation angles of the fissioning fragments in the reaction plane (their moment of inertia and angular momenta are $\Im_{1,2}$ and $L_{1,2}$, respectively), $a_{1,2}=\frac{R}{2}\pm\frac{\left(R_{1}-R_{2}\right)}{2}$ is the distance from the center of the fragments to the middle point of neck region, and $R_{1,2}$ is the fragment radii. The symbol $R$ is distance between the fragment centers.
The total angular momentum $L=\ell+L_{1}+L_{2}$ is preserved. The symbol $\mu_{R}$ is reduced mass, and $\gamma_{tan}$ is the tangential friction force of the colliding nuclei.
Here, it is called sliding friction.
The phenomenological nuclear friction forces for separated nuclei are expressed in terms of $\gamma_{tan}^{F}$ for sliding friction using the Woods-Saxon radial form factor described in Refs. [6, 7].
The sliding friction are described as $\gamma_{tan}=\gamma_{t}^{0}F\left(\zeta\right)$, where the radial form factor $\ F\left(\zeta\right)=\left(1+\exp^{\zeta}\right)^{-1}, \zeta=\frac{\left(\xi-\rho_{F}\right)}{a_{F}}$. $\gamma_{t}^{0}$ denote the strength of the tangential friction, respectively. $\rho_{F} \sim$ 2 fm and $a_{F} \sim$ 0.6 fm are the model parameters, and $\xi$ is the distance between the nuclear surfaces $\xi=R-R_{contact}$, where $R_{contact}=R_{1}+R_{2}$ [6].
The symbols separated by $m_{ij}$ and $\gamma_{ij}$ stand for the shape-dependent collective inertia and friction tensors elements, respectively.
We adoped the hydrodynamic inertia tensor $m_{ij}$ in Werner-Wheeler approximation for the velocity field [16]. The normalized random force $R_{i}\left(t\right)$ is assumed to be white noise: $\langle R_{i} (t) \rangle=0$ and $\langle R_{i} (t_{1})R_{j} (t_{2})\rangle = 2 \delta_{ij}\delta (t_{1}-t{2})$.
According to Einstein relation, the strength of the random force $g_{ij}$ is given $\gamma_{ij}T=\sum_{k}{g_{ij}g_{jk}}$, where $T$ is the temperature of the compound nucleus calculated from the intrinsic energy of the composite system.
The adiabatic potential energy is defined as
\begin{eqnarray}
&V_\mathrm{{adiab}}\left(q,L,T\right)=V_{LD}\left(q\right)+
\frac{\hbar^{2}L\left(L+1\right)}{2\mathcal{I}\left(q\right)} \nonumber
+V_{SH}\left(q,T\right), \nonumber \\
&V_{LD}\left(q\right)=E_{S}\left(q\right)+E_{C}\left(q\right), \nonumber \\
&V_{SH}\left(q,T\right)=E_{shell}^{0}\left(q\right)\Phi\left(T\right), \nonumber \\
&\Phi\left(T\right)=\exp \left(-\frac{E^{\ast}}{E_{d}}\right).
\end{eqnarray}
Here, $\mathcal{I}\left(q\right)$ represents the moment of inertia of the rigid body with deformation $q$.
The centrifugal energy generated from the angular momentum $L$ of the rigid body is also taken into account.
$V_{LD}$ and $V_{SH}$ are the potential energy of the finite range liquid drop model and the shell correction energy taking into account of the temperature dependence, respectively.
The symbol $E_{shell}^{0}$ indicates the shell correction energy at $T=0$.
The temperature dependent factor $\Phi\left(T\right)$ is explained in Ref. [14], where $E^{\ast}$ indicates the excitation energy of the compound nucleus. $E^{\ast}$ is given $E^{\ast}=aT^{2}$, where $a$ is the level density parameter.
The shell damping energy $E_{d}$ is selected as 20 MeV. This value is given by Ignatyuk et al. [17].
The symbols $E_{S}$ and $E_{C}$ stand for generalized surface energy [18] and Coulomb energy, respectively.
\section*{Results and discussion}
\subsection*{Fragment mass and total kinetic energy}

\quad The effect of the tangential friction with the present framework is described below. Figure 1 shows the mass distribution and the average total kinetic energy ($\overline{TKE}$) for different strength of tangential friction in the reaction system of $^{48}$Ca+$^{248}$U.

\begin{figure}[h]
\centering
\includegraphics[width=0.75\textwidth]{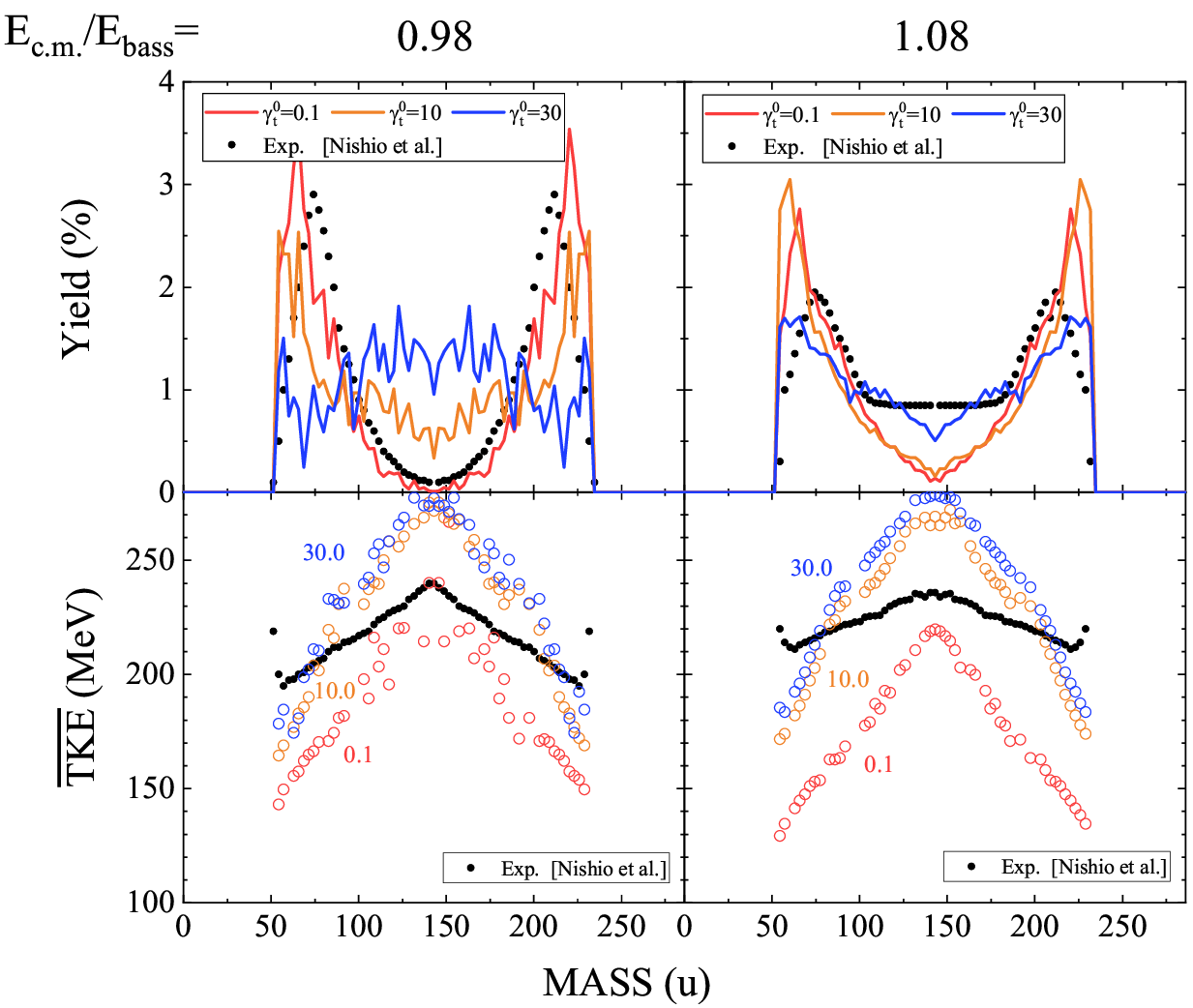}\\
\scriptsize{Figure 1. Fission fragment mass distribution and average total kinetic energy ($\overline{TKE}$) in the reaction of $^{48}$Ca+$^{238}$U at the different strength of tangential friction for the different $E_{c.m.}/E_{bass}$ values. The calculation results are shown in the mass region of $54 \le A \le 232$ by the colored lines in the upper panels and by the open circles in the lower panels.  The experimental fission yields and the $\overline{TKE}$ are shown by the dots [19].}
\label{fig:1}
\end{figure}
As the stronger tangential friction is assumed, the fission yields increase in the mass symmetric region. And consequently, $\overline{TKE}$ tends to increase. At the low incident energy shown in the upper left panel of Figure 1, mass distribution drastically changes from the asymmetric fission to the symmetric one as increasing the strength of tangential friction.  At high energies, the dominant fission mode does not change dramatically.
The calculation result of the mass distribution at $E_ {c.m.}=190$MeV ($E^{*}=30.9$MeV) assuming the value of $\gamma_t^0=0.1$ MeV s fm$^{-2}$ is good agreement with the experimental value. But $\overline{TKE}$ is underestimated in this case.
On the other hand, the mass distribution at $E_{c.m.}=210$MeV ($E^{*}=50.9$MeV) can be reproduced the tendency of the experimental value with the friction parameter $\gamma_t^0=30.0$ MeV s fm$^{-2}$. However, $\overline{TKE}$ shows considerably higher   than the experimental value even if in this strong friction case. The reason why $\overline{TKE}$ is greatly affected with $\gamma_t^0$ is in the different deformation which concerns with the Coulomb energy at scission as presented in the followings.
\subsection*{The behaviors of scission point}

\begin{figure}[h]
\centering
\includegraphics[width=0.7\textwidth]{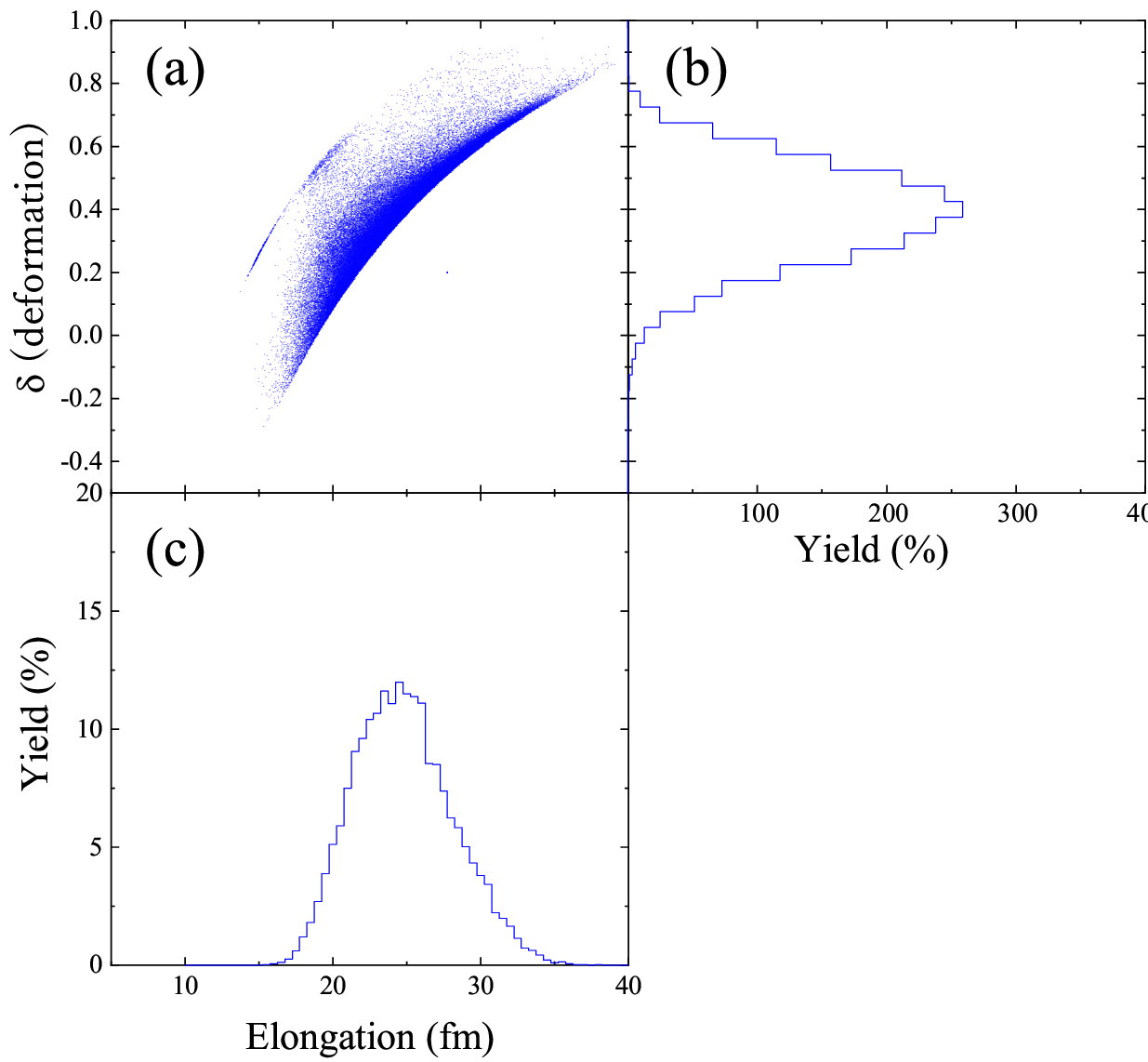}\\
\scriptsize{Figure 2. The scission points distribution, the $\delta$ distribution, and the distribution of the distance between centre of mass of right and heavy fragments for $\gamma_t^0=0.1$ MeV s fm$^{-2}$ and $E_ {c.m.}/E_{bass}=0.98$.}
\label{fig:2}
\end{figure}
\begin{figure}[h]
\centering
\includegraphics[width=0.7\textwidth]{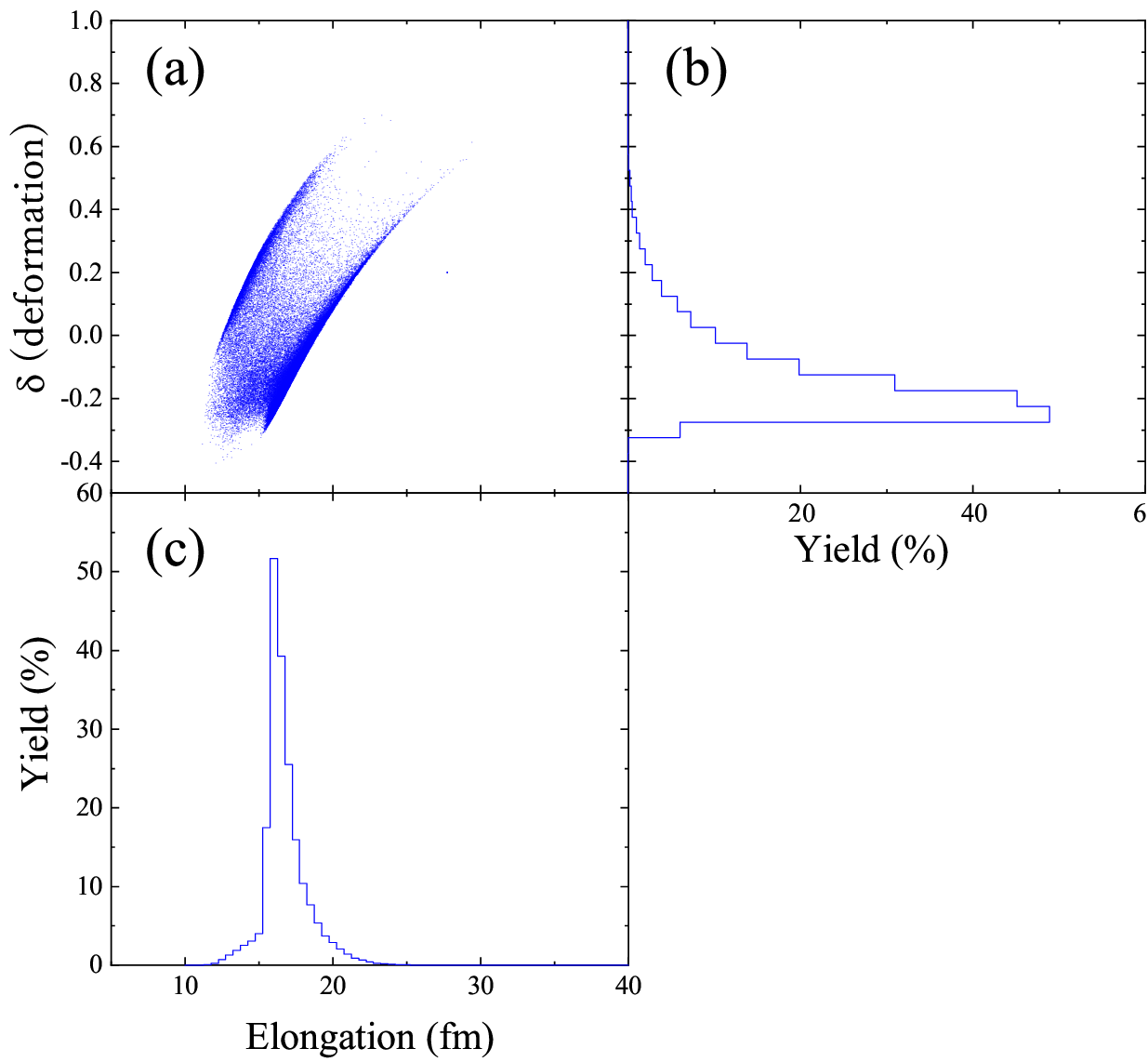}\\
\scriptsize{Figure 3. Same as Fig. 2 but for $\gamma_t^0=30.0$ MeV s fm$^{-2}$ and $E_{c.m.}/E_{bass}=1.08$.}
\label{fig:3}
\end{figure}

\quad Figure 2(a)-(c) show the scission points distribution, the $\delta$ distribution, and the distance distribution between centre of mass of colliding system with $\gamma_t^0=0.1$ MeV s fm$^{-2}$, respectively. From Fig. 2(a), it can be seen that the scission points distribute around $\delta \sim 0.3$ and the elongation of about 25fm.  These calculations show that the fissioning nuclei separate in the considerably stretched shape.
Further, the investigation is made on the behavior of scission point under the strong friction.  Figures 3(a), (b), and (c) show the same distributions for $\gamma_t^0=30.0$ MeV s fm$^{-2}$.
Figure 3(a) shows that the scission points are more populated in the lower left than that in the case of Fig. 2(a).
The peak of the distribution locates on the $-\delta$ region as shown in Fig. 3(b), and the scission occurs at about the elongation of 16fm from Fig. 3(c).
If $\gamma_t^0$ increases, the deformation parameter $\delta$ moves toward negative region and the fissioning fragments separate with a small distance between their centers. This means that the fission occurs in the compact configuration.
The TKE is greatly affected because the nuclear shape at scission depends on the strength of tangential friction. The TKE is expressed as follows:
\begin{eqnarray}
&TKE=PKE+PCE=PKE+e^2\frac{Z_{1}Z_{2}}{d_{sci}},
\end{eqnarray}
where PKE and PCE denote the pre-scission kinetic energy and the pre-scission Coulomb energy, respectively. $e^{2}=1.44$MeV fm, $Z_{1}, Z_{2}$ are the charge of each fragment, $d_{sci}$ is the distance between centre of mass of light and heavy parts of the nucleus at the scission point.
From Fig. 2, for the small $\gamma_t^0$ the scission distance $d_ {sci}$ shifts to the large value. Thus, the TKE value becomes lower than the experimental one. On the other hand, for large $\gamma_t^0$, the value of $d_{sci}$ shows too small. Therefore, it is considered that the TKE value is higher than the experimental value. The reason why the calculated TKE value is not good agreement with experimental data is that the TKE could not be evaluated accurately, because the amount of the conversion of the internal energy into PKE can not be easily speculated and the behaviour of the scission point depends largely on the strength of tangential friction.

\subsection*{The fusion mechanism and the tangential friction}
\quad The compound nucleus (CN) are formed after the projectile nucleus is captured by the target nucleus. The capture occurs by overcoming the coulomb force of the colliding system. Here, we investigate the dependence of capture cross section on the tangential friction. The capture cross sections in the $^{48}$Ca+$^{238}$U system are show in Fig. 4 as the function of $\gamma_t^0$ in the range of $E_{c.m.}/E_{bass}$=0.930-1.186.
\begin{figure}[h]
\centering
\includegraphics[width=1.0\textwidth]{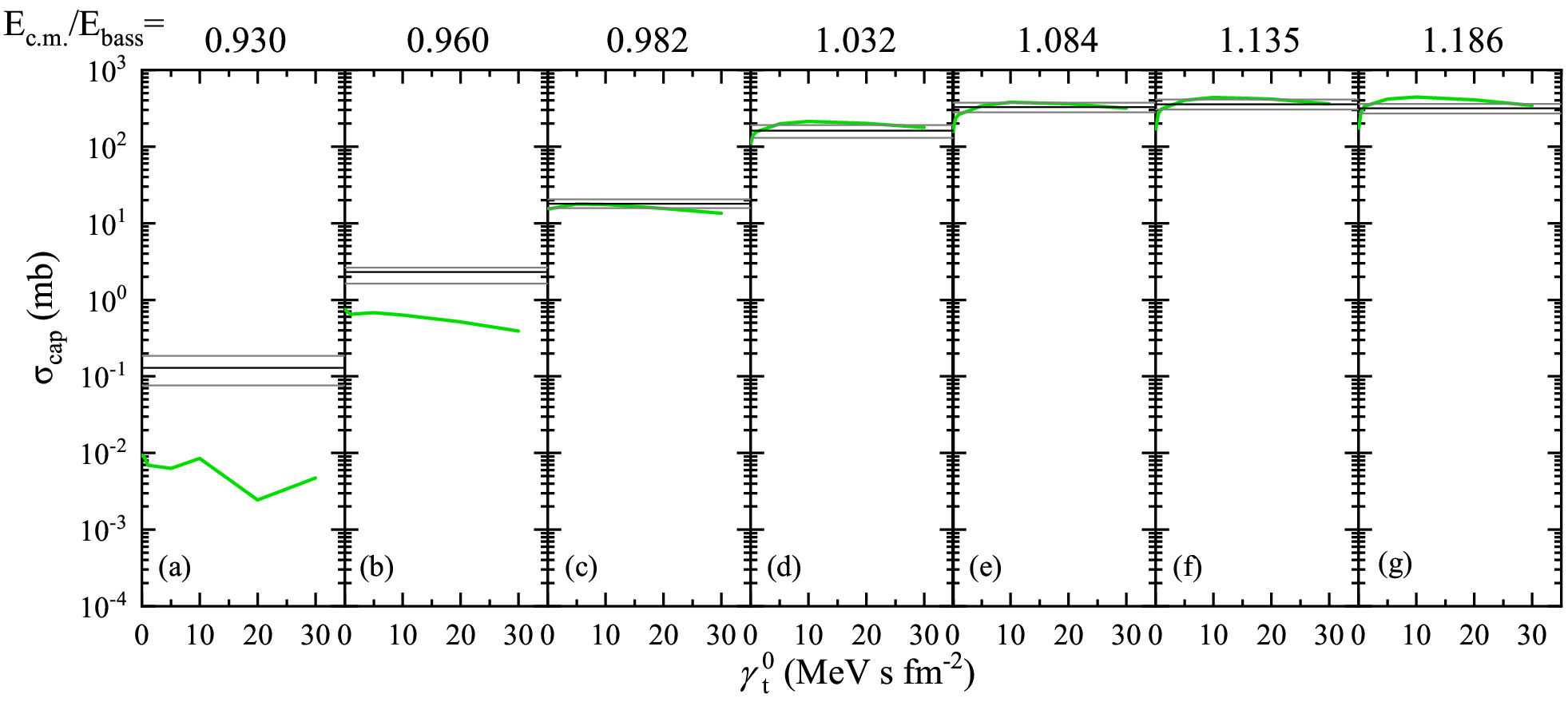}\\
\scriptsize{Figure 4. (a)-(g) The $\gamma_t^0$ dependence of the capture cross section $\sigma_{cap}$ obtained by the Langevin calculation in the range of $E_{c.m.}/E_{bass}$=0.930-1.186 in the reaction system $^{48}$Ca+$^{238}$U. The horizontal black and gray line show experimental data and its error bar, respectively [19].}
\label{fig:4}
\end{figure}
\begin{figure}[p]
\begin{tabular}{cc}
\begin{minipage}[t]{1.0\hsize}
\centering
\includegraphics[width=0.8\textwidth]{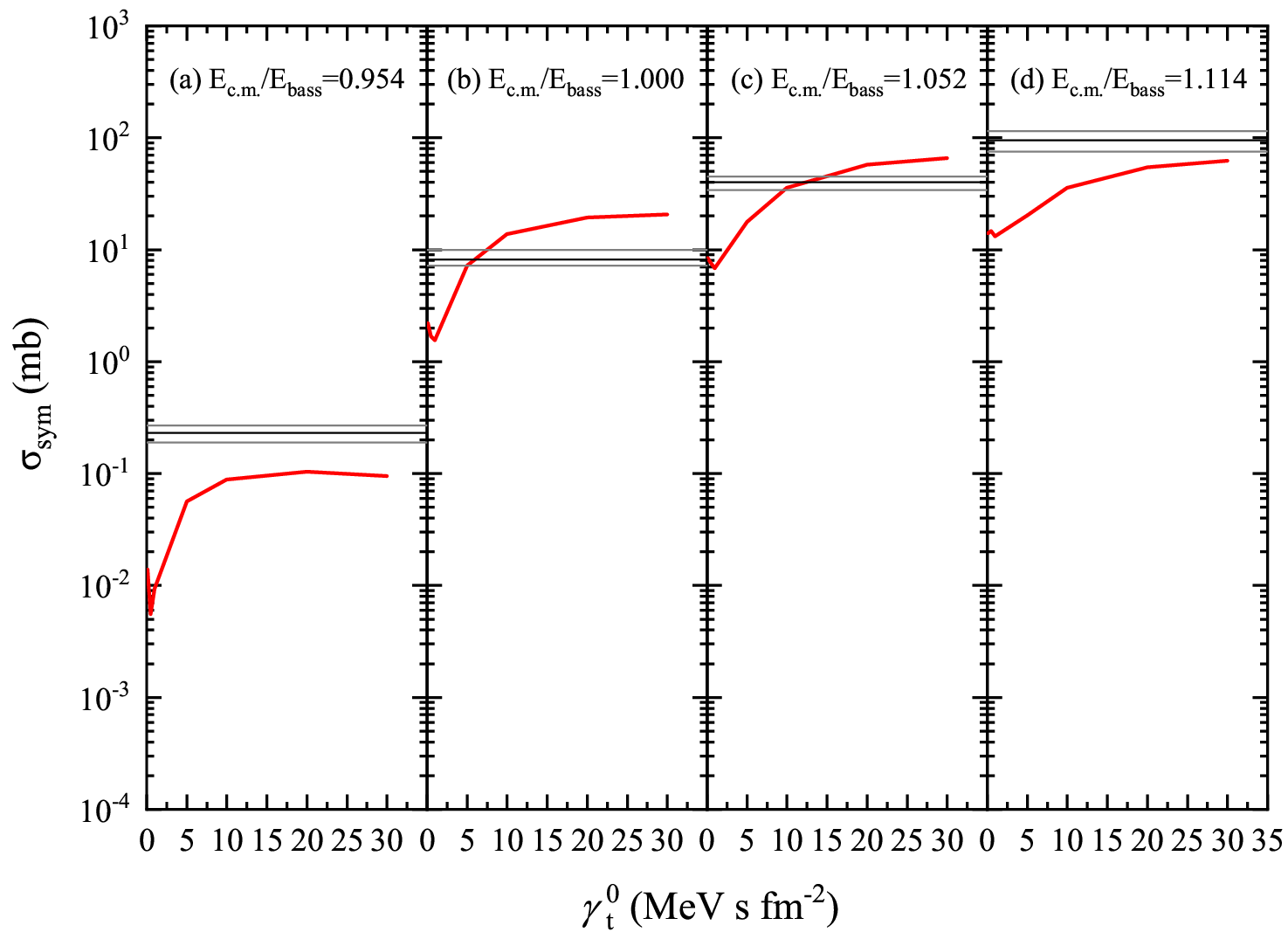}\\
\scriptsize{Figure 5. The $\gamma_t^0$ dependence of the fission cross section $\sigma_{sym}$ defined as $A_{CN}/2 \pm 20$u at (a) $E_{c.m.}/E_{bass}$ = 0.954, (b) $E_{c.m.}/E_{bass}$ = 1.000 and (c) $E_{c.m.}/E_{bass}$ = 1.052 and (d) $E_{c.m.}/E_{bass}$ = 1.114, in the  reaction $^{48}$Ca+$^{238}$U. The horizontal black and gray line show experimental data and its error bar, respectively [20].}
\label{fig:5}
\end{minipage} \\\
\begin{minipage}[t]{1.0\hsize}
\centering
\includegraphics[width=0.8\textwidth]{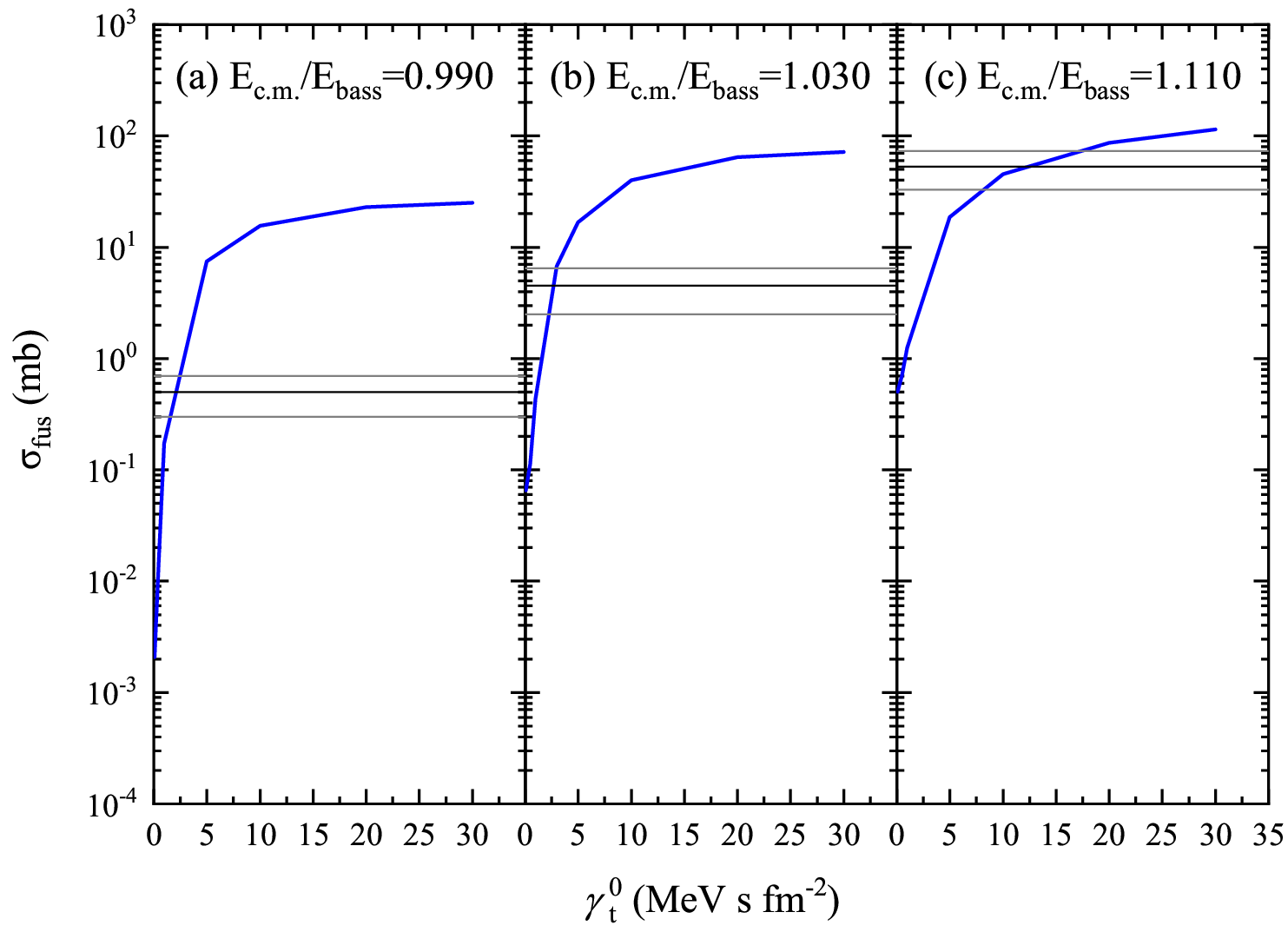}\\
\scriptsize{Figure 6. The $\gamma_t^0$ dependence of the section $\sigma_{fus}$ for (a) $E_{c.m.}/E_{bass}$ = 0.990, (b) $E_{c.m.}/E_{bass}$ = 1.030 and (c) $E_{c.m.}/E_{bass}$ = 1.114, in the  reaction $^{48}$Ca+$^{238}$U. The horizontal black and gray line show experimental data and its error bar, respectively [21].}
\label{fig:6}
\end{minipage}
\end{tabular}
\end{figure}
The horizontal black and the gray line show experimental data and its error bar, respectively [19]. Though the capture cross section depends on the incident energy, it is almost constant with the variation of $\gamma_t^0$, because the property of the radial form factor of the sliding friction discussed in the previous section.
The calculation results are not in good agreement with the experiments under the barrier energy due to being out of application of the Langevin-type approach.
To form CN, it is also necessary the mass drift occurring towards mass-symmetry. Here, We investigate the relationship between the tangential friction and the mass drifts towards mass-symmetry. Figure. 5 shows the $\sigma_{sym}$ defined $A_{CN}/2 \pm 20$u in the range of $E_{c.m.}/E_{bass}$ = 0.954-1.114 for $^{48}$Ca+$^{238}$U system.
The horizontal black and the gray line show experimental data and its error bar, respectively [20].
It is possible to infer the adjustable value of $\gamma_t^0$ for $\sigma_{sym}$ above the $E_{bass}$ from the figure.
 However, under the $E_{bass}$ our calculation underestimates irrelevantly to the $\gamma_t^0$, because our model is also unapplicable in this energy region.

Next, we investigate the effect of the tangential friction on the fusion cross section. The fusion cross section $\sigma_{fus}$ is estimated by the summation of events whose trajectories enter the fusion box [14]. The definition of fusion box is $\lbrace|\alpha|<0.3,\ \delta<-0.5z+0.5\rbrace$ in the present calculation.
Fig. 6(a)-(c) show the fusion cross section at the different energies as the function of tangential friction $\gamma_t^0$.
The horizontal black line and the gray line show experimental data and its error bar, respectively [21].
The fusion cross section increases in proportion to the strength of friction because the trajectories can enter the fusion box by lowering the fusion barrier due to the dissipation by the tangential friction.
At the lower energies, the variation in fusion cross section owing to the tangential friction is larger than that at the higher energies.
This is because the fusion hindrance due to centrifugal energy appears prominently at the incident energy near the barrier, but at the energy sufficiently higher than the barrier it is easy to form a compound nucleus independent of the friction.
At $E_{c.m.}/E_{bass}$ = 0.990, the experimental value can be reproduced with a small value of $\gamma_t^0$ around 1 MeV s fm$^{-2}$ as shown in Fig. 5.
On the other hand, in higher incident energy, the experimental value can be fitted by about one order higher value of  when the value of $\gamma_t^0$.

Here, the $\gamma_t^0$ determined from Fig. 4, Fig. 5 and Fig. 6 as a function of $E_{c.m.}/E_{bass}$ show together in Fig. 7 with the experimental data [19, 20, 21].
\begin{figure}[h]
\centering
\includegraphics[width=0.7\textwidth]{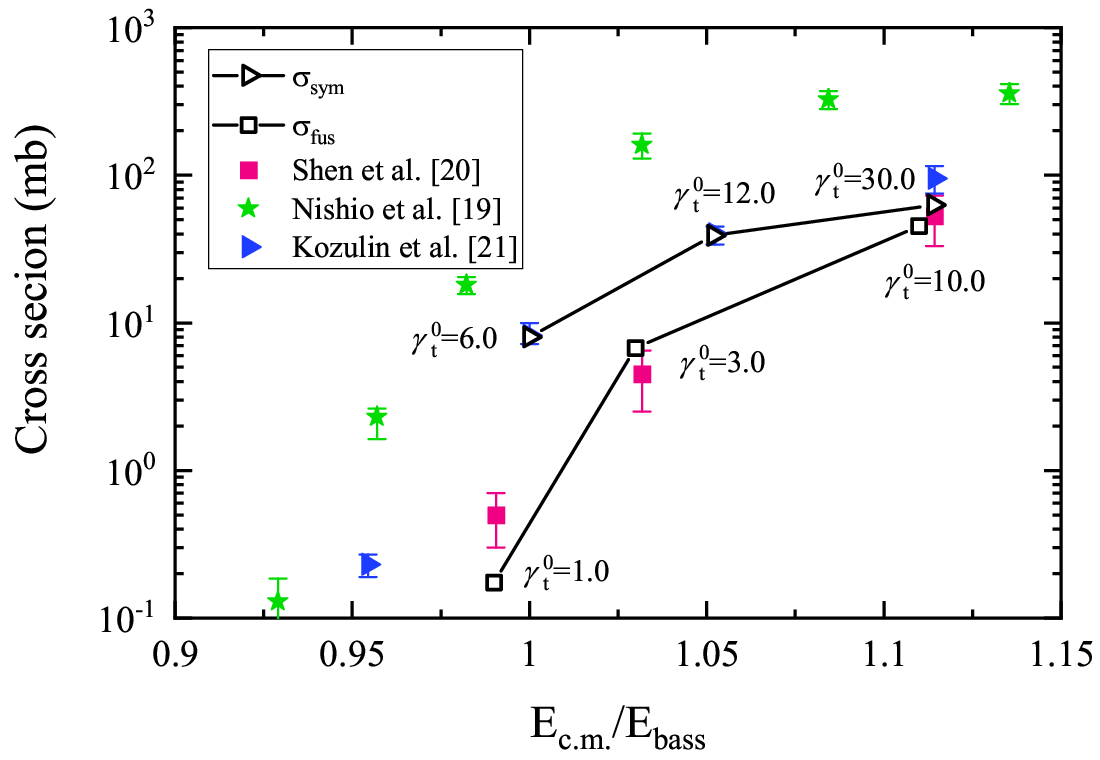}\\
\scriptsize{Figure 7. The mass-symmetric fission (open black triangles) and the fusion cross sections (open black squares) for each $\gamma_t^0$ determined using Fig. 4, Fig. 5 and Fig. 6. The experimental cross section for forming CN (filled squares [21]) and that for full momentum transfer (FMT) fusion (stars [19]), and the yield of the symmetric fragments with masses $A_{CN}/2 \pm $20u (filled triangles [20]) are plotted.
}
\label{fig:7}
\end{figure}
\begin{figure}[h]
\centering
\includegraphics[width=0.7\textwidth]{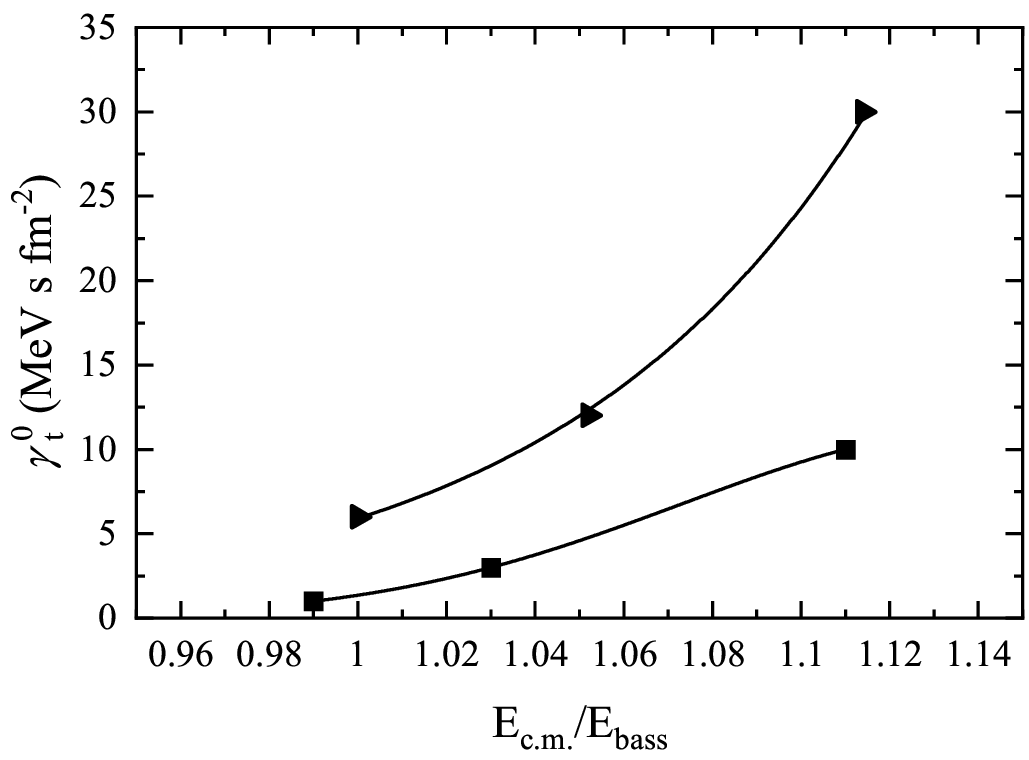}\\
\scriptsize{Figure 8. The values of $\gamma_t^0$ in Fig.7 are plotted as the function of $E_{c.m.}/E_{bass}$. Filled black triangles and squares show $\gamma_t^0$ value determined by comparing calculation results with experimental of the cross section of $A_{CN}/2 \pm $20u [20] and fusion cross section [21], respectively.
}
\label{fig:8}
\end{figure}
The experimental data are shown as square points mean the cross section for the compound nucleus (CN). The symbol stars in the figure contain both quasi-fission (QF) and fusion-fission (FF) events, so they are considered to be the capture cross section. The cross section for triangles is mass-symmetric fission cross section.
The calculation results for $\sigma_{sym}$ and $\sigma_{fus}$ are in good agreement with the experimental data by taking each $\gamma_t^0$ value indicated in the figure.
%
%

Summarizing the tangential friction determined from the two kinds of experimental data,
the values of $\gamma_t^0$ are plotted in Fig.~8 as a function of $E_{c.m.}/E_{bass}$. Filled black triangles and squares show $\gamma_t^0$ value determined from the comparison with the cross section of $A_{CN}/2 \pm $20u [20] and the fusion cross section[21], respectively.

The value of $\gamma_t^0$ tends to increase with the $E_{c.m.}/E_{bass}$. The variation of $\gamma_t^0$ determined from $\sigma_{fus}$ is milder than one determined from $\sigma_{sym}$ with respect to the $E_{c.m.}/E_{bass}$.
This is because fusion reactions relate with the dissipative effects strongly in the process from the contact point to the stage of sufficient mass drift attained.
In the high energies, the contact surface interacts strongly (the increase of the  tangential friction) due to increasing the relative velocity of the projectile-target.
It can be seen that the tangential friction greatly affects the enhancement of fusion.

\section*{Conclusions}
\quad The dissipative effect is quite delicate for fusion process, especially in the reaction process in superheavy element production. In this paper, we presented how the tangential friction affects the fusion cross section and the TKE of fission fragments. The value of $\gamma_t^0$ which can reproduce the experimental fusion cross section was clarified. 
The value of $\gamma_t^0$ are extracted from both the experimental data of fusion $\sigma_{fus}$ and the yield of symmetric like fragments $\sigma_{sym}, A_{CN}/2\pm20u$.
It is shown that the compatible value of $\gamma_t^0$ with the experiments increases with the increase of the incident energy in proportion to the power of the relative velocity between the projectile an the target.
Figure 8 means that the larger tangential friction is needed to reproduce the quasi-fission data, comparing with the case of complete fusion data.
These results indicate how the mass drift to form the CN and the composite system for quasi-fission depends on the tangential friction in the different way corresponding to the different region of angular momentum induced in the entrance channel.
The result is useful for the prediction to the fusion cross section in the unknown superheavy elements.
On the one hand, it is noted that the relation between the strength of $\gamma_t^0$and the experimental $\overline{TKE}$ could not be resolved because of the difficulty in the estimation on the thermal energy conversion into the pre-scission kinetic energy in the process from the saddle to the scission point.


\section*{Acknowledgments }
The Langevin calculations were performed using the cluster computer system (Kindai-VOSTOK) which is supported by JSPS KAKENHI Grant Number 20K04003.

\begin{flushleft}

[1] E. Margaret Burbidge, et al., Rev. Mod. Phys {\bf 29} (1957) 547. \href{https://doi.org/10.1103/RevModPhys.29.547} {[\textit{\textcolor[rgb]{0,0,1}{CrossRef}}]}

[2] P. Indelicato, et al., Nature {\bf 498}, (2013) 40-41. \href{https://doi.org/10.1038/498040a} {[\textit{\textcolor[rgb]{0,0,1}{CrossRef}}]}

[3] W. Nazarewicz, Nat. Phys. {\bf 14}, (2018) 537-541. \href{https://doi.org/10.1038/s41567-018-0163-3} {[\textit{\textcolor[rgb]{0,0,1}{CrossRef}}]}

[4] R. Bass, Nuclear reactions with heavy ions, (1980) 321-323. \href{https://link.springer.com/gp/book/9783540096115} {[\textit{\textcolor[rgb]{0,0,1}{CrossRef}}]}

[5] D.H.E. Gross et al., Phys. Rep. C {\bf 45} (1978) 175. \href{https://doi.org/10.1016/0370-1573(78)90031-5} {[\textit{\textcolor[rgb]{0,0,1}{CrossRef}}]}

[6] V.~Zagrebaev and W.~Greiner, J. Phys. G {\bf 31}, (2005) 825-844. \href{https://iopscience.iop.org/article/10.1088/0954-3899/31/7/024/meta} {[\textit{\textcolor[rgb]{0,0,1}{CrossRef}}]}

[7] V.~Zagrebaev and W.~Greiner, J. Phys. G {\bf 34}, (2007) 2265-2277. \href{https://iopscience.iop.org/article/10.1088/0954-3899/34/11/004/meta} {[\textit{\textcolor[rgb]{0,0,1}{CrossRef}}]}

[8] V.I.~Zagrebaev et al., Phys. Part. Nuclei {\bf 38}, (2007) 469. \href{https://doi.org/10.1103/PhysRevC.83.044618} {[\textit{\textcolor[rgb]{0,0,1}{CrossRef}}]}

[9] G.F.~Bertsch, Z. Phys. A {\bf 289}, (1978) 103. \href{https://doi.org/10.1007/BF01408501} {[\textit{\textcolor[rgb]{0,0,1}{CrossRef}}]}

[10] W. Cassing and W. N\"{o}renberg, Nucl. Phys. A {\bf 401}, (1983) 467. \href{https://doi.org/10.1016/0375-9474(83)90361-5} {[\textit{\textcolor[rgb]{0,0,1}{CrossRef}}]}

[11] A. Diaz-Torres, Phys. Rev. C {\bf 69}, (2004) 021603. \href{https://doi.org/10.1103/PhysRevC.69.021603} {[\textit{\textcolor[rgb]{0,0,1}{CrossRef}}]}

[12] J.~Maruhn and W.~Greiner, Z. Phys {\bf251}, (1972) 431. \href{https://doi.org/10.1007/BF01391737} {[\textit{\textcolor[rgb]{0,0,1}{CrossRef}}]}

[13] K.~Sato, A.~Iwamoto, K.~Harada, S.~Yamaji and S.~Yoshida, Z. Phys. A {\bf288}, (1978) 383. \href{https://doi.org/10.1007/BF01417722} {[\textit{\textcolor[rgb]{0,0,1}{CrossRef}}]}

[14] Y.~Aritomo and M.~Ohta, Nucl. Phys. A {\bf744}, (2004) 3-14. \href{https://doi.org/10.1016/j.nuclphysa.2004.08.009} {[\textit{\textcolor[rgb]{0,0,1}{CrossRef}}]}

[15] Y.~Aritomo, Phys. Rev. C {\bf80}, (2009) 064604. \href{https://doi.org/10.1103/PhysRevC.80.064604} {[\textit{\textcolor[rgb]{0,0,1}{CrossRef}}]}

[16] K.T.R~Davies, A.J.~Sierk and J.R.~Nix, Phys. Rev. C {\bf13},  (1976) 2385. \href{https://doi.org/10.1103/PhysRevC.13.2385} {[\textit{\textcolor[rgb]{0,0,1}{CrossRef}}]}

[17] A.N.~Ignatyuk, G.N.~Smirenkin and A.S.~Tishin, Sov. J. Nucl. Phys {\bf21}, (1975) 255. \href{https://inis.iaea.org/search/search.aspx?orig_q=RN:6208426} {[\textit{\textcolor[rgb]{0,0,1}{CrossRef}}]}

[18] H.J.~Krappe, J.R.~Nix and A.J.~Sierk, Phys. Rev. C {\bf20},  (1979) 992. \href{https://doi.org/10.1103/PhysRevC.20.992} {[\textit{\textcolor[rgb]{0,0,1}{CrossRef}}]}

[19] K. Nishio et al., Phys. Rev. C {\bf 86} (2012) 034608. \href{https://doi.org/10.1103/PhysRevC.86.034608} {[\textit{\textcolor[rgb]{0,0,1}{CrossRef}}]}

[20] E.M. Kozulin et al., Phys. Rev. C {\bf99} (2019) 014616. \href{https://doi.org/10.1103/PhysRevC.99.014616} {[\textit{\textcolor[rgb]{0,0,1}{CrossRef}}]}

[21] W.Q.~Shen et al., Phys. Rev, C {\bf 36} (1987) 115. \href{https://doi.org/10.1103/PhysRevC.36.115} {[\textit{\textcolor[rgb]{0,0,1}{CrossRef}}]}

\end{flushleft}

\end{document}